\documentclass[11pt,twoside]{article}

\usepackage{asp2004}
\usepackage{epsf}
\usepackage{psfig}
\usepackage{lscape}
\usepackage{graphicx}

\begin{document}
\title{Turbulence Effects at Small Scales}   
\author{A. Beresnyak and A. Lazarian }   
\affil{Astronomy Department, University of Wisconsin-Madison, 475 Charter St., Madison,
WI 53705}    

\begin{abstract} 
It is most natural to assume that mysterious Small Ionized and Neutral Structures (SINS) in
diffuse ISM arise from turbulence. There are two obvious problem with
such an explanation, however.
First of all, it is generally believed that at the small scales turbulence should be damped.
Second, turbulence with Kolmogorov spectrum cannot be the responsible
for the SINS.
We consider, however, effects, that provide spectral index flatter
than the Kolmogorov one and allow action at very
small scales. These are the shocks that arise in high Mach number turbulence
and transfer of energy to
small scales by instabilities in cosmic rays. Our examples indicate that the origin of
SINS may be discovered through systematic studies of astrophysical turbulence.   
\end{abstract}


\section{Turbulence and SINS}   

Various observations covered well in this volume
indicate the existence of
structure in the neutral and ionized ISM on AU spatial scales, causing
concern about how well we understand the basics of the ISM
physics. In the straightforward interpretation, these clouds should be
extremely overdense and and overpressured. Evaporation arguments point
to very short lifetimes of such clouds, yet the AU-scale structures
in the diffuse medium seem quite common. 
In this volume the reader can find many articles dealing with the observed properties of
SINS. Therefore we do not dwell upon this issue. Instead we adopt a broad theoretical
approach and ask a question whether turbulence can produce enough activity on small scales
to be considered as a potential source of SINS.

The attractiveness of turbulence as an origin of SINS is that it can produce a generic universal
explanation. Indeed, turbulence is really ubiquitous in astrophysics, including the
ISM and circumstellar regions. However, the most popular Kolmogorov turbulence
model can not account for SINS. Kolmogorov turbulence has a spectrum of $E(k)\sim k^{-5/3}$,
the fluctuations of density at scale $k\sim 1/l$ arisen from advection by turbulence
are $(\delta \rho_k)^2\sim k E(k) \sim k^{-2/3}$, i.e. they decrease with the scale. SINS, on the
contrary, are connected to rather large density contrasts at small scales. This would violate
the assumption of Kolmogorov turbulence that it is only weakly compressive at large outer scale.
Fortunately, astrophysical turbulence is not limited to the Kolmogorov spectrum. If the
spectrum of density fluctuations scales as $E(k)\sim k^{-n}$, $n<1$, then $(\delta \rho_k)^2$
increases with the decrease of turbulence scale.   

As we see from the example above, in terms of turbulence, the origin of SINS is a quantitative
question related to the spectra of turbulent motions. This question cannot be dealt using brute
force of computers. Indeed, from the point
of view of fluid mechanics astrophysical turbulence 
is characterized by huge Reynolds numbers, $Re$, which is  
the inverse ratio of the
eddy turnover time of a parcel of gas to the time required for viscous
forces to slow it appreciably. For $Re\gg 100$ we expect gas to be
turbulent and this is exactly what we observe in HI (for HI $Re\sim 10^8$).
In fact, very high astrophysical $Re$ and its magnetic counterpart
magnetic Reynolds number $Rm$
 (that can be as high as $Rm\sim 10^{16}$) present a big problem for numerical simulations
that cannot possibly get even close to the astrophysically-motivated numbers\footnote{
This caused serious concerns that while present codes can produce simulations that
resemble observations, whether numerical simulations reproduce reality well 
(see McKee 1999, Shu et al. 2004).}. 
The currently available 3D
simulations can have $Re$ and $Rm$ up to $\sim 10^4$.  Both scale as
the size of the box to the first power, while the computational effort
increases as the fourth power (3 coordinates + time), it is not
feasible to resolve the actual interstellar turbulent cascade to test whether
SINS somehow emerge at the end of it. Another approach based on (a) establishing scaling
relations that can be tested with the modern computers and (b) testing these relations
with observational data seems promissing. In what follows we discuss the part (a). Part
(b) is addressed in a recent review by Lazarian (2006). 
 
\section{Density spectrum in supersonic MHD turbulence}

A very important insight into the
incompressible MHD turbulence by Goldreich \& Shridhar (1995) (henceforth GS95)
has been followed by progress in understanding of compressible MHD turbulence
(Lithwick \& Goldreich 2001, Cho \& Lazarian 2003, henceforth CL03, 
Cho, Lazarian \& Vishniac 2003).
In particular, simulation in Cho \& Lazarian (2003) showed that Alfv\'enic cascade evolves
on its own\footnote{
The expression proposed and tested in CL03 shows that the coupling of Alfv\'enic and
compressible modes is appreciable at the injection scale if the injection velocity
is comparable with the {\it total} Mach number of the turbulence, i.e. with
$(V_A^2+C_S^2)^{1/2}$, where $V_A$ and $C_S$ are the Alfv\'en and sound velocities
respectively. However, the coupling gets marginal at smaller scales  as
turbulence cascades and
turbulent velocities get smaller.}
and it exhibits Kolmogorov type scaling (i.e. $E\sim k^{-5/3}$)
and scale-dependent
anisotropy of the Goldreich-Shridhar type (i.e. $k_{\|}\sim k_{\perp}^{2/3}$)
even for high Mach number turbulence. While slow modes exhibit similar
scalings and anisotropy, fast modes show isotropy. The density scaling
obtained in Cho \& Lazarian (2003) was somewhat puzzling. At low Mach
numbers it was similar to slow modes, while it got isotropic 
for high Mach numbers.

Density power spectrum shallower than the Kolmogorov
one was reported in a number of observations (Deshpande, Dwarakanath \& Goss 2000, 
Padoan et al 2003) and the relation between the SINs and the 
shallow power spectrum was advocated in Deshpande (2000). 

For high Mach numbers we expect shocks to develop; the density should be perturbed
most dramatically by those shocks. Naively, one would think that density perturbations
will have a random-shock spectrum of $k^{-2}$ which leaves very little perturbation
on small scales. Also, in sub-Alfv\'enic turbulence magnetic field is dynamically
important, so we expect to observe some anisotropy which is typical for magnetic turbulence.
None of these two properties is actually observed in simulations. On the contrary,
the spectrum was typically rather shallow and there was no significant anisotropy.
This mystery is resolved as follows: shocks in isothermal fluid can have very large 
density contrasts, up to sonic Mach squared, however the conservation of mass and
positive sign of density allows only small regions with high-density or clumps.
These clumps totally dominate the spectrum of density. Being close to delta-functions,
they generate rather shallow spectrum. Being distributed randomly in space, they mask
any anisotropy originally present due to Alfv\'enic shearing.

Our calculations in Beresnyak, Lazarian \& Cho (2006, henceforth BLC06) 
demonstrated that the spectrum of 
density for high Mach number MHD turbulence is shallow (see Fig.~1). This potentially
may be important for SINs, although we do not attempt to provide quantitative arguments
for this at the present stage.  
The rms deviation of density for a subsonic case is consistent with
prediction $M^2$ for low beta (CL03), and the rms deviation of log-density
for supersonic case is around unity regardless of a Mach number. The distributions
are notably broader for higher Mach numbers, though.
 
\begin{figure}
\includegraphics[width=0.495\textwidth]{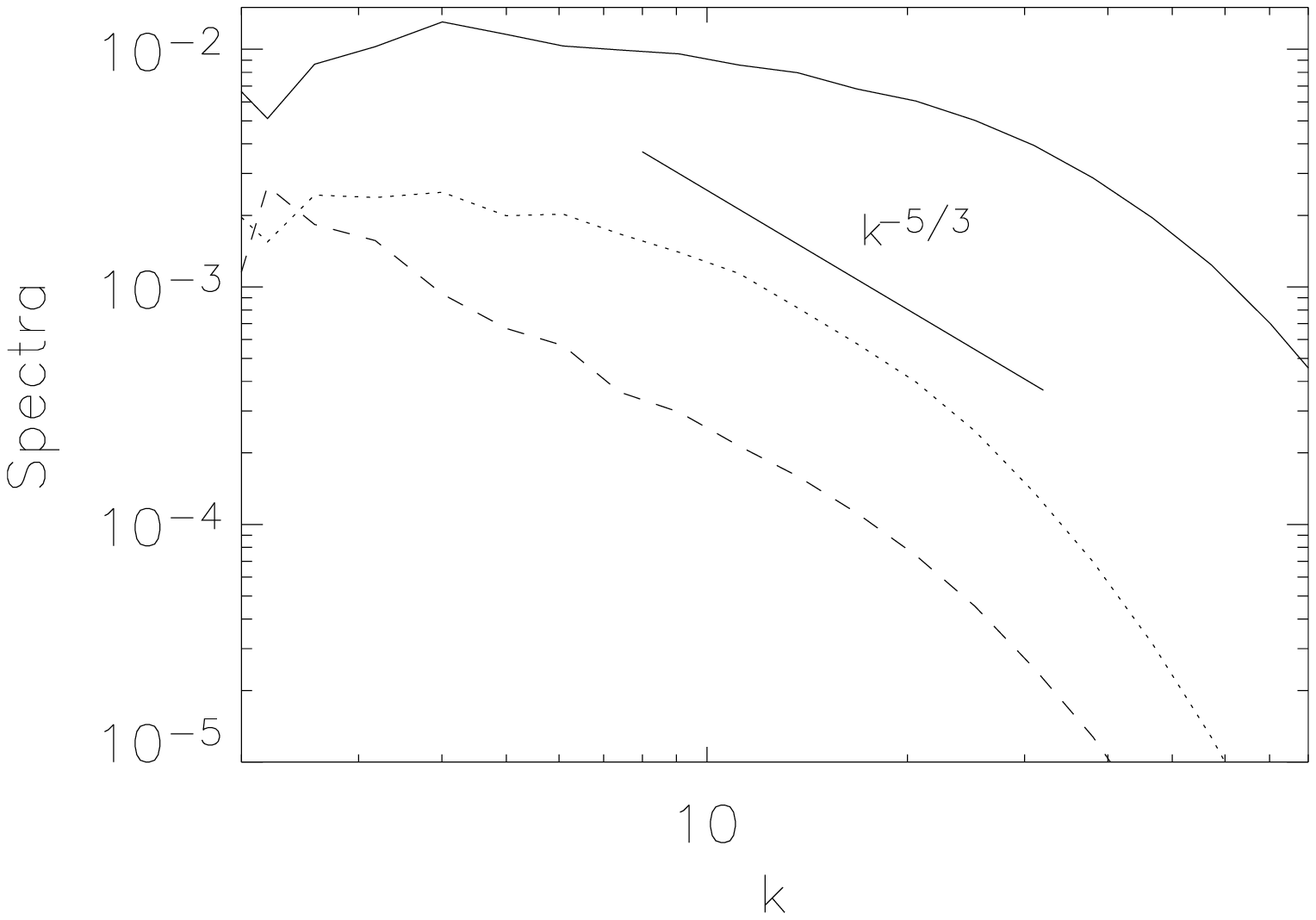}
  \hfill
  \includegraphics[width=0.495\textwidth]{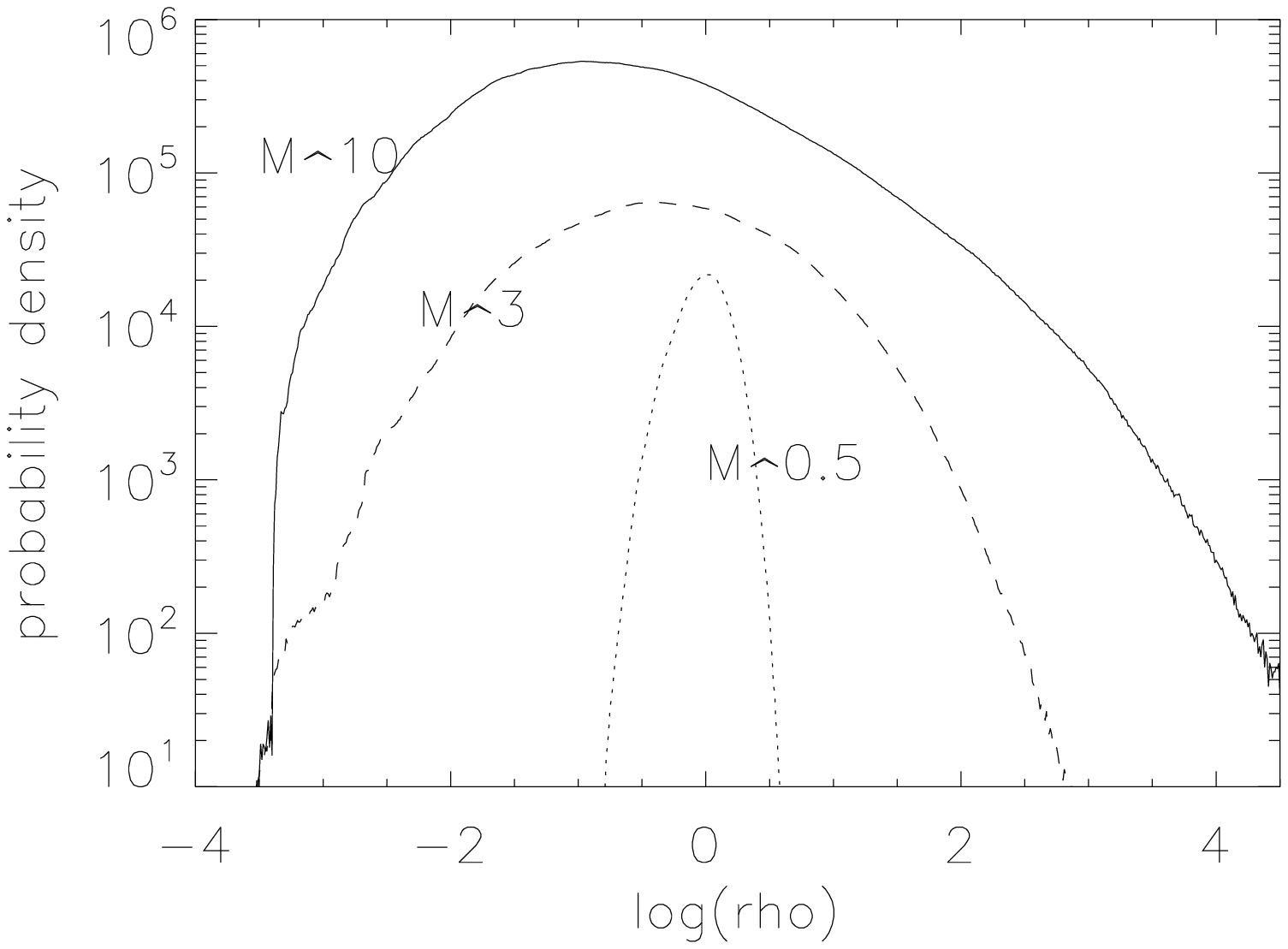}
 \caption{{\it Left panel}: Mach number is 10, power spectra
 of: {\it solid} -- density, {\it dashed} -- velocity, {\it dotted}
-- logarithm of density. {\it Right Panel}: Probability density function for a density in 
 simulation with Alfv\'enic Mach number around unity and various sonic
Mach numbers.}
\label{spectrum}
\end{figure}
Dimension of the high-density structures is between 1 and 2, being
viewed as a flatted filaments or elongated pancakes. There were no evident
preferred orientation of these structures along or perpendicular to
the mean magnetic field. Maximum density value in a Mach 10
data cube was around $3\times10^3\rho_0$. We also noted in BLC06 that the
randomly distributed high-density clumps 
suppress any anisotropy originating from motions at small scales.

In magnetically dominated medium that we deal with it is reasonable
to assume that the corresponding shocks move material along magnetic
field lines the same way that the slow modes do in subsonic case.
The shocks are randomly oriented and therefore the clumpy structure
that we observe does not reveal any noticeable anisotropy.
In fact, the density perturbations associated with such shocks should not 
be correlated with the magnetic field strength enhancement, which
our data analysis confirms. Note, that 
due to different
reasons, this correlation also weak for density fluctuations induced by slow modes
low Mach number MHD turbulence (see CL03).

Interestingly enough, similar results in terms of shallow density spectrum
are seen in pure high Mach number hydrodynamic simulations
(Kim \& Ryu 2005). This reinforces the idea that the primary mechanism for
generating of flat spectrum of density is the random multiplication/division
of density in supersonic flows which, together with mass conservation, form
isolated high peaks of density. Does this mean that magnetic field is totally
unimportant for the structure of density in supersonic turbulence? Not at all
-- we were able to reveal the underlying density structure, that comes from
Alfv\'enic shearing by using a log-density instead of density for spectra
and structure functions. The structure function of log-density have shown
very familiar Goldreich-Sridhar scale-dependent anisotropy, typical for
almost any quantity in sub-Alfv\'enic compressible turbulence (see Fig. 2).

\begin{figure}
  \includegraphics[width=0.29\textwidth]{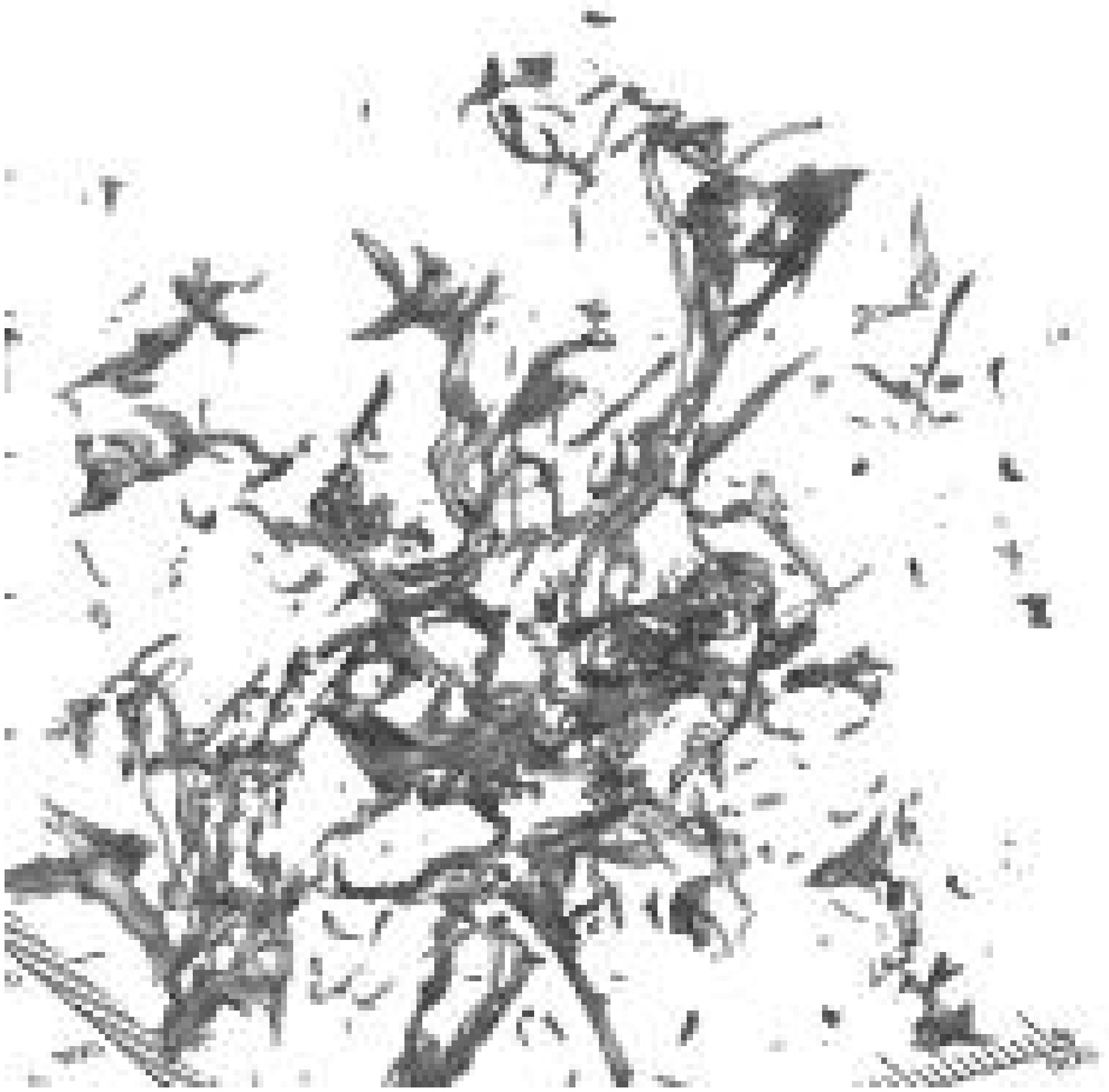}
  \hfill
  \includegraphics[width=0.29\textwidth]{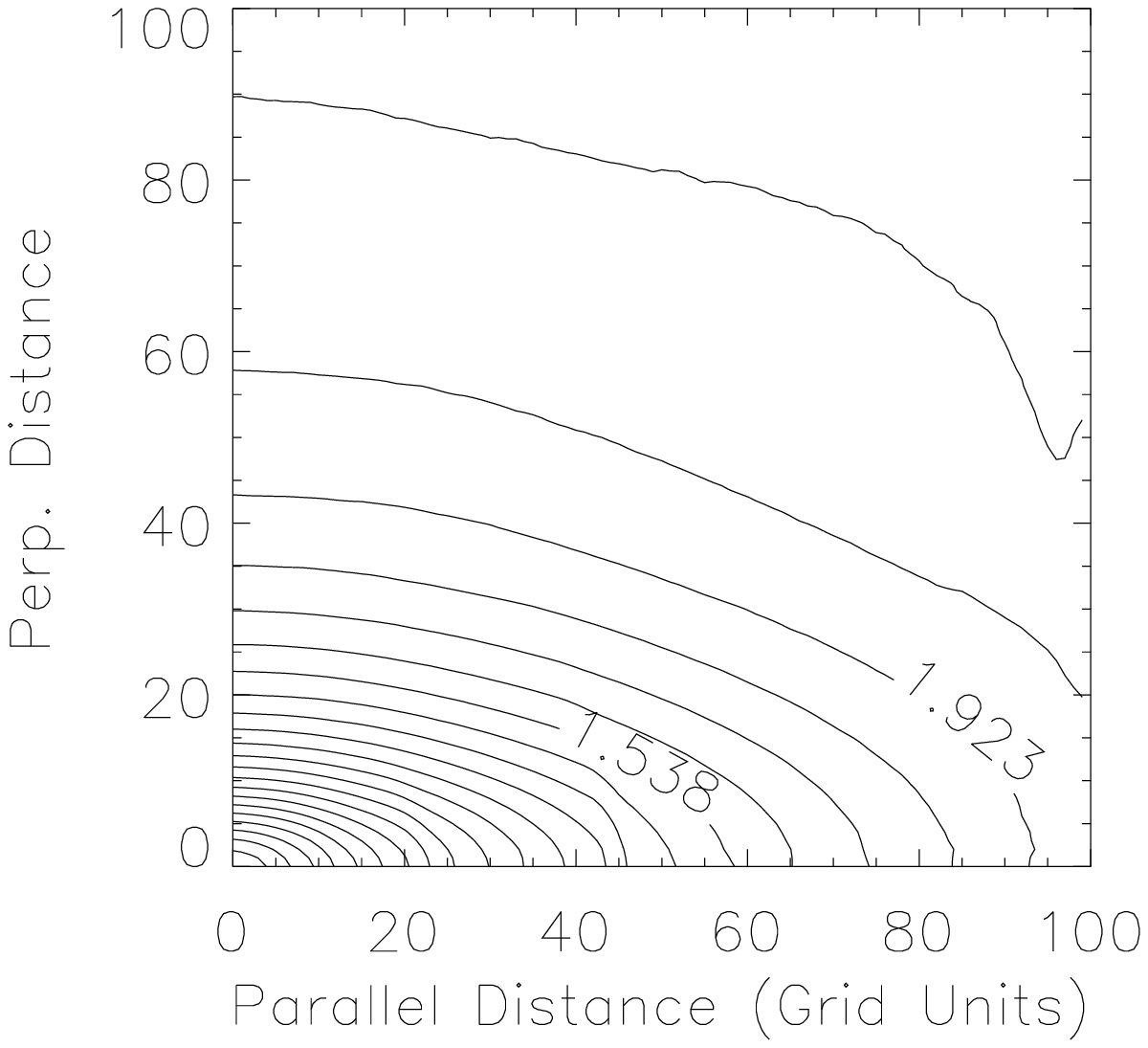}
  \hfill
  \includegraphics[width=0.38\textwidth]{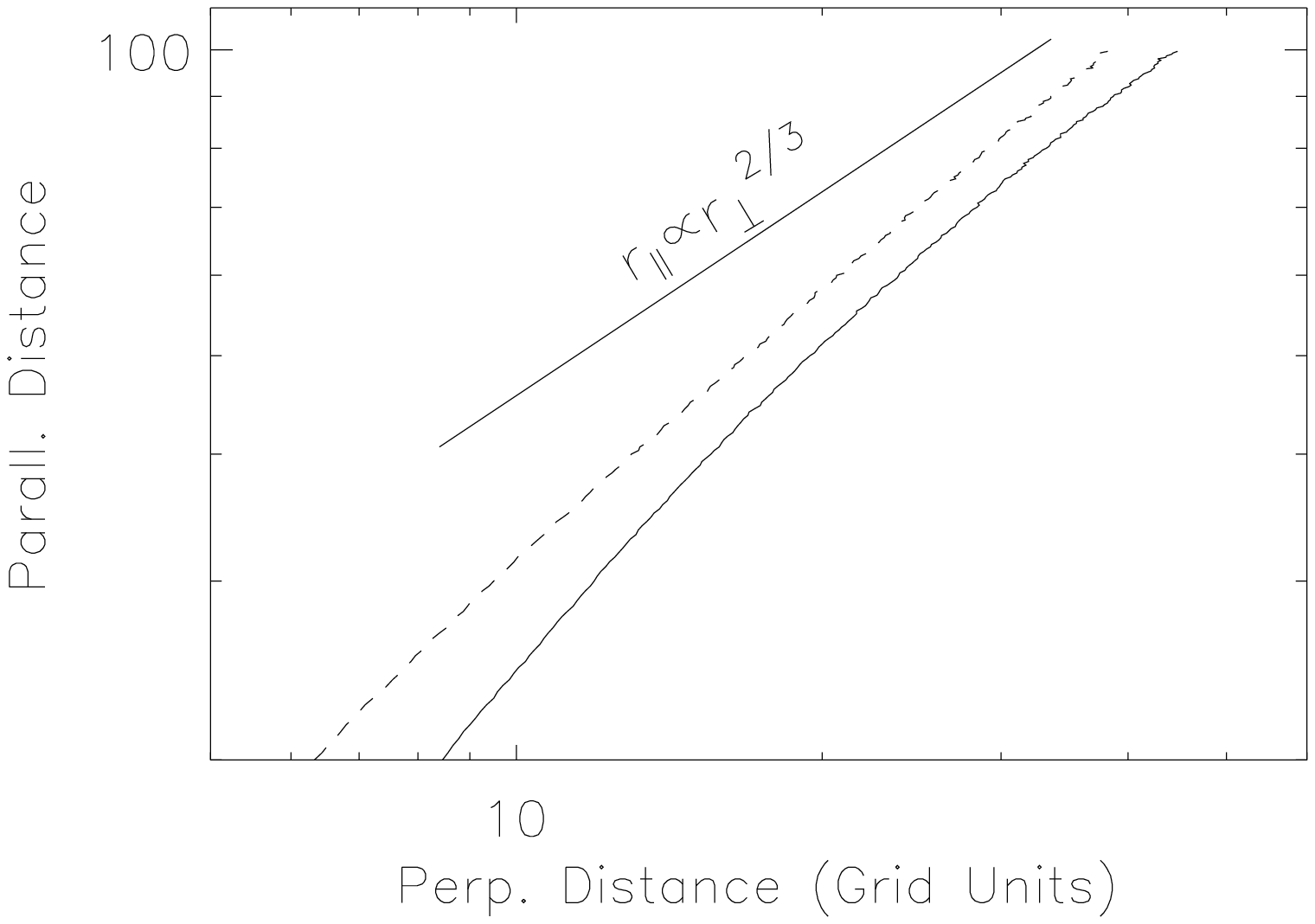}
  \caption{ $M_s\sim 10$, {\it Left Panel}: The isosurfaces of density, corresponding to 10
 mean densities.  {\it Central Panel}: Contours of equal correlations
of log-density, calculated respective to the local magnetic field.
They reveal pronounced anisotropy of density structure.
{\it Right panel}: $r_\|$ and $r_\perp$ for anisotropic eddies revealed by
logarithm of density, and comparison with GS95 law, {\it solid} $M_s\sim 10$, {\it dashed} $M_s\sim 3$. }
\label{SF}
\end{figure}

Calculations of the density scaling with a more extended sample of simulations with a variety
of orginary and Alfven Mach numbers in Kowal, Lazarian \& Beresnyak (2006) are consistent with
our findings above. The explanation of the shallow spectrum of density presented above
suggests that if we include gravity the clumps will be denser and the power spectrum of density
will be flatter or could be even rising ($\delta$-functions will give
the 1D power spectrum of $k^2$)

In terms of the relation to SINs, we may note, that cooling may make
interstellar gas more pliable to compression than the isothermal gas that
we used in the simulations. This, could result in more density contrast when
the original gas is warm. This calls for detailed studies that include cooling.

\section{Alfvenic Slab motions induced by Cosmic Rays}

In this section we describe a new mechanism to transfer turbulent
energy to small scales bypassing the usual turbulent cascade. This
mechanism, described in Lazarian \& Beresnyak (2006) allow us to have
relatively large-amplitude slab-type Alfvenic perturbations on small scales.
Even though MHD Alfven waves do not perturb density, small-scale kinetic
Alfven waves (KAW) do. Also the slab Alfven waves in a compressible media
is subject to the well-known parametric instability that produces density
perturbations.

The close connection 
between the cosmic ray (CR) power-law and the turbulence power-law is the point
that was stressed by Randy Jokipii in many of his presentations (see Jokipii 2001).
The physical essence
of this relation and, in fact, the very understanding of the interaction of
CR and turbulence are currently the issues of intensive research.
It is well known that the propagation of CR depends on the scaling of
the ISM turbulence. Most of the studies in the field use isotropic or
slab-like Alfvenic turbulence the origin of which does not follow from either
theoretical models of MHD turbulence
 (see Goldreich \& Shridhar 1995; review by Cho \& Lazarian 2005 and
references therein) or numerical simulations of MHD turbulence
 (see Cho \& Lazarian 2003 and references therein).
On the contrary, the interactions of 
Alfvenic turbulence with scale dependent anisotropy that is
consistent with both theoretical predictions and numerics
were shown to be inefficient for cosmic ray scattering
(Chandran 2000, Yan \& Lazarian 2002). Using the scalings for fast
modes obtained in Cho \& Lazarian (2003), Yan \& Lazarian (2002)
identified fast modes as the principal scattering agent for standard
models of MHD turbulence, provided that the energy is injected at scales larger
than a pc. In this model that follows from this work the differences
in damping of fast modes result in the differences in CR propagation in
different ISM phases. This induces substantial changes on the models of 
CR propagation in our galaxy (Yan \& Lazarian 2004) and in clusters of
galaxies (Brunetti \& Lazarian 2006). 

While the above changes of the picture of CR propagation and possibly acceleration are
inevitable in the quasi-linear models of scattering by MHD turbulence injected at large scales,
in an astrophysically realistic situation with the pressure of CRs close to thermal
or magnetic pressure, it is reasonable to ask whether the feedback of CRs to
MHD turbulence is important. This question was addressed in Lazarian \& Beresnyak (2006).
CRs react very differently from ordinary gas when the magnetized fluid is
compressed, if the scale of compression is less than CR mean free
path.  Such scale compressions or expansions change only the component
of CR momentum that is perpendicular to magnetic field (due to the so-called
``adiabatic invariant'' conservation), while the component 
of the momentum parallel to magnetic field stays the same.
Such state is, however, unstable. 

We calculated the rate of CR kinetic instability that
transfers energy to small scales, creating small-scale Alfvenic
perturbations that are not a part of the global MHD cascade. These
perturbations are more like waves moving along magnetic field lines
and thus are very different from highly anisotropic GS95 Alfvenic modes.

Quantitatively, we consider a power-law distribution of CRs
$F_0 \sim p^{-\alpha-2}$ where $\alpha$ is
conveniently defined as the power-law index for a one-dimensional
distribution (or particle density). For example, around the Earth $\alpha
\sim 2.6$ up to the energies of $10^{14}$~eV. 
The the growth rate of the
cosmic-ray-Alfv\'en gyroresonance instability (henceforth GI) can be
estimated as:
\begin{equation}
\gamma_{\rm CR}(k_{\|}) = \pm \omega_{pi} \frac{n_{\rm CR}(p>m\omega_B/k_{\|})}{n}AQ,
\label{stream}
\end{equation}
where $n_{\rm CR}(p>m\Omega/k_\|)$ is the number density of CRs with
momentum larger than the minimal resonant momentum for a wave vector
value of $k_\|$, $m$ is the proton mass, $n$ is the density of the thermal
plasma, $\omega_{pi}$ is the ion plasma frequency. $Q$ is a numerical
factor that depends on the index of cosmic rays $\alpha$.  The $\pm$ sign corresponds to the
two MHD modes. We shall concentrate on the Alfv\'en mode, corresponding to
the plus sign, as those are less subjected to linear damping. 
We show that when anisotropy is
created by compressive turbulence, the anisotropy factor
$A=(p_\perp-p_\|)/p_\|$
is small and changes its sign on the scale of the mean free path,
depending on two competitive mechanisms -- scattering which tends to
isotropize momentum distribution, and magnetic field compression
which tends to make it oblate or prolate. 

Assuming that anisotropy arises from the turbulent
compressions with amplitude $\delta v$ at large scale, the
factor $A$ is equal to $2\delta v/v_A$, where $v_A$ is the Alfven velocity.
Then the expression for the instability rate can be written as
\begin{equation}
\gamma_{\rm CR}(r_p)=\frac{\delta v}{L_i}\left(\frac{r_p}{r_0}\right)^{-\alpha +1},
\label{instab_main}
\end{equation}
where $r_p$ is a Larmor radius of a CR resonant with a particular wave
vector $k_\|=m\Omega/p$, $r_0$ is the 1~GeV proton Larmor radius and
\begin{equation} 
L_i=3.7\cdot10^{-7}\frac{1}{Q} 
\left(\frac{B}{5\cdot10^{-6}~\mbox{G}}\right)
\left(\frac{4\cdot10^{-10}~\mbox{cm}^{-3}}{n_{\rm CR}(r_p>r_0)}\right)~{\rm pc}.
\label{lili}
\end{equation}
This CR instability gets energy from external turbulent compressions of
CR  at the mean free path scale and directly transfers it to scale of the CR 
Larmor radius. 

There are two non-linear processes limit the growth rate of the instability. First of all,
the magnetic perturbations generated at the Larmor radius of CR backreact on CR by limiting
the mean free path $\lambda$. If the
change of magnetic field direction is $\phi\sim \delta B/B$ the
scattering that is a random walk requires $N\sim 1/\phi^2$ interaction
and
\begin{equation}
\lambda\sim Nr_p\sim r_p/\phi^2\sim r_p B^2/(\delta B)^2,
\label{mean_path}
\end{equation}
where we designated $\delta B$ as the magnetic field perturbation
pertaining to a particular wavenumber, i.e.  $\delta B^2\approx
E(k)k$. We can consider $\delta B$ as a function of either $k$ or the
resonant Larmor radius $r_p$ (see Longair 1994). $\delta B$ grows as the instability
grows, which in turn reduces $\lambda$. 
On the other
hand, it is the mean free path $\lambda$ which determines the scale at
which compressions of the magnetic field are important. This can be
understood as follows: the CR distribution ``remembers'' the perturbed
value of the magnetic field and its anisotropy only during the
time the typical particle travels its mean free path. Once particles
scatter significantly, the anisotropy of the distribution is
effectively ``reset''. As a result only low amplitude motions on
scales less than $\lambda$ excite the instability which limits
the degree of anisotropy $A$ attainable.
 
The instability grows as $d(\delta B^2)/dt=\gamma_{\rm CR} (\delta B^2)$
where the injection of energy is happening at the scale of the mean
free path, i.e.
\begin{equation}
\gamma_{\rm CR}\approx\frac{v_{A}}{L_i}
\left(\frac{r_p}{L}\right)^{\mu}\left(\frac{\delta B}{B}\right)^{-2\mu}
\left(\frac{r_p}{r_0}\right)^{-\alpha+1},
\label{instab_limit}
\end{equation}
where eqs. (\ref{instab_main}) and (\ref{mean_path}) were used.  We see
that according to the above equations $\delta B$ perturbations will grow
as $t^{1/2\mu}$ thus reducing $\lambda$ virtually to $r_p$.
If not for other processes, it is possible
to show that this suppression would reduce $\lambda$ to virtually the CR gyroradius,
inducing Bohm-type diffusion of CRs.

However, the collection of waves with different wavelengths created by the instability is
also subject to steepening with the rate of $\gamma_{\rm steep} \approx -(\delta B/B)^2 k_{\|} v_A$,
which combined with the effect of limiting $\lambda$ above provides:
\begin{equation}
\frac{\delta B}{B}\approx
\frac{r_0^{1/2}}{L_i^{1/(2\mu+2)} L_{\phantom{i}}^{\mu/(2\mu+2)}}
\left(\frac{r_p}{r_0}\right)^{(\mu-\alpha+2)/(2\mu+2)}.
\label{amplitude}
\end{equation}
For $\alpha=2.6$ and $\mu=1/3$ Eq.~(\ref{amplitude}) produces a rather shallow spectrum
of perturbations, $E(k)\approx (\delta B)^2/k\sim k^{-0.8}$.

The calculations above are valid provided that $\lambda$ is larger than
the compressive mode cutoff scale $l_{\rm cut}$. If, on the other hand,
$l_{\rm cut}>\lambda$ the compression for the instability
is supplied from the eddies at the damping scale, namely,
$\delta v/v_A\sim (l_{\rm cut}/L)^{1/3}(\lambda/l_{\rm cut})$.
In this case instead of eq. (\ref{amplitude}) one gets
\begin{equation}
\frac{\delta B}{B}\approx\left(\frac{r_0^{1/2}}{L_i^{1/4}L_{\phantom{i}}^{\mu/4}l_{\rm cut}^{(1-\mu)/4}}\right)^{1/4}
\left(\frac{r_p}{r_0}\right)^{(3-\alpha)/4},
\label{amplitude2}
\end{equation}
which, for the same value of $\alpha=2.6$, corresponds to a steeper 
spectrum of $E(k)\sim k^{-1.2}$.

\begin{figure}
  \includegraphics[width=0.25\textwidth]{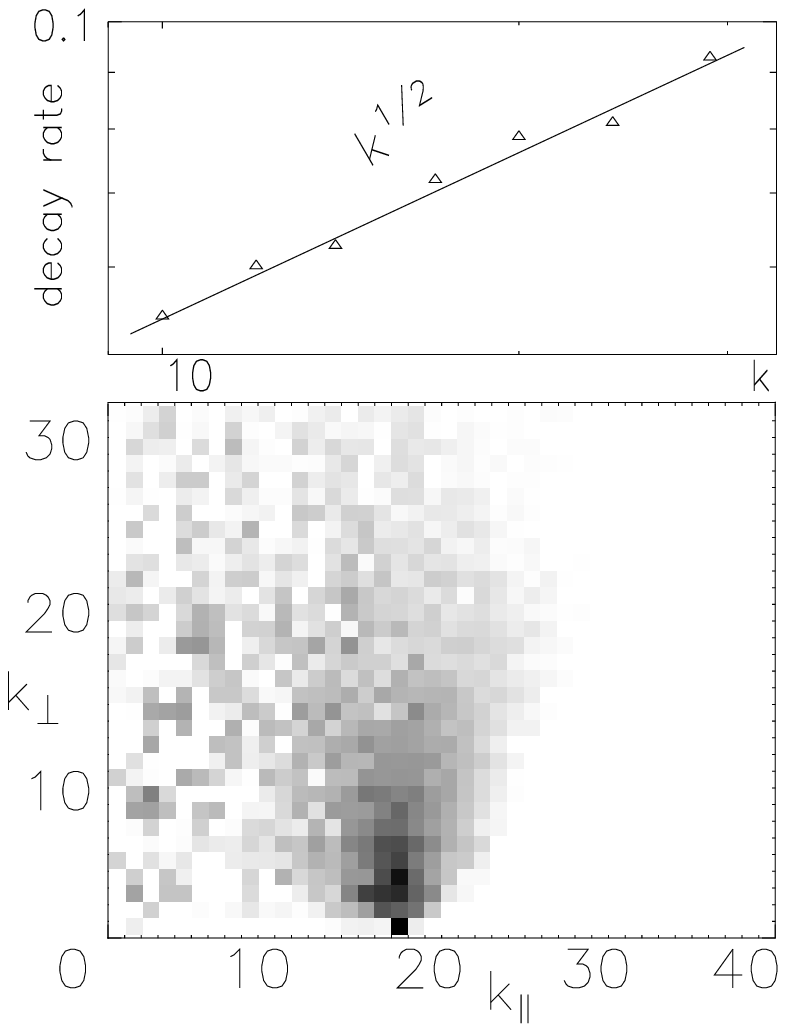}
  \hfill
  \includegraphics[width=0.50\textwidth]{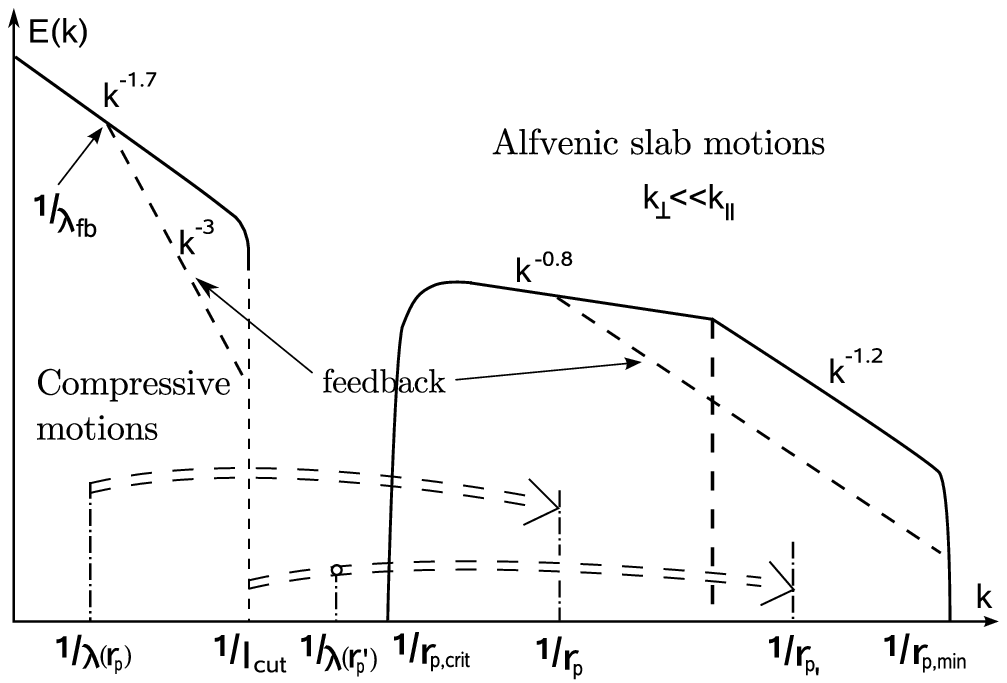}
  \caption{ {\it Left Panel}: Decorrelation of a plane, $k_\perp=0$ Alfv\'en wave by
  turbulence. Lower picture shows the energy density of a wave in
  cylindrical k-space. Alfv\'en waves were injected at $k_{\|}=17$.
  Wave energy is being transferred in the direction of $k_\perp$ axis,
  which is typical for decorrelation by MHD turbulence. 
  Upper plot shows decay rate of the wave vs its wavenumber. 
{\it Right Panel}: Energy density of compressive modes and Alfv\'enic slab-type
waves, induced by CRs. The energy is transferred from
the mean free path scale to the CR Larmor radius scale.
If the mean free path falls below compressive motions cutoff or feedback
suppression scale, the spectrum of slab waves becomes steeper. See more on feedback
in Lazarian \& Beresnyak (2006).
}
\label{SF}
\end{figure}

The ambient Alfvenic turbulence provides yet another source of non-linear damping
of the waves generated by the instability. This process is analogous to the suppression of
the streaming instability (Yan \& Lazarian 2002, Farmer \& Goldreich 2004).
Our numerical testings of the scalings of the corresponding damping rate, i.e. 
\begin{equation}
\gamma_{\rm turb}\sim -k_{\bot} v_{\bot} \sim -v_A k_\perp^{2/3} L^{-1/3} 
\sim - v_A r_p^{-1/2} L^{-1/2},
\label{turb}
\end{equation}
are shown in Figure~3 (left). The encouraging agreement of Eq.~\ref{turb} with the results
of MHD simulations makes us confident in using this rate, which up to the sign convention,
coincides with the prediction in Farmer \& Goldreich (2004). Combining 
eqs.~(\ref{instab_limit}), (\ref{amplitude}) and (\ref{turb}) we find
that for $\alpha>5/3$ our instability is damped for all scales, larger than
\begin{equation}
r_{p,{\rm crit}}\approx r_0\left(L_{\phantom{i}}^{1-\mu}r_0^{\mu+1}L_i^{-2}\right)^{1/(2\alpha-\mu-3)}.
\label{crit_rad}
\end{equation}
Therefore the spectrum of plane Alfv\'en waves given by
eq.~(\ref{amplitude}) will protrude from $r_{p,{\rm crit}}$ down to
$r_{p,{\rm min}}$ which corresponds to minimum energies of CRs.

\section{Summary}

Astrophysicists always expect to see Kolmogorov turbulence. These expectations
may not be met in reality.
We presented two examples of astrophysical 
processes that provide distinctly non-Kolmogorov spectra. One 
case is the density in supersonic MHD turbulence, another is the velocity and magnetic field
perturbations at small scales. This work is suggestive that a possible explanation of the
SINS should be sought invoking non-Kolmogorov turbulence.


\acknowledgements 
AB thanks
IceCube project for support of his research. AL acknowledges the NSF grant
AST-0307869 and the support from the Center for Magnetic
Self-Organization in Laboratory and Astrophysical Plasmas. 


\end{document}